\documentclass[aps,prd,reprint,twocolumn,nofootinbib]{revtex4-2}
\pdfoutput=1 
\usepackage{graphicx}
\usepackage{dcolumn}
\usepackage{bm}
\usepackage{color}
\usepackage{textcomp}
\usepackage{subfigure}
\usepackage{feynmp}
\usepackage{slashed}
\usepackage{amsmath,amssymb,array}
\usepackage{enumerate}
\usepackage{hhline}
\usepackage{hyperref}

\hypersetup{colorlinks=true,urlcolor=[rgb]{0,0,0.5},citecolor=[rgb]{0,0,0.5},linkcolor=[rgb]{0,0,0.5}}
\newcommand{\beqa}{\begin{eqnarray}}
\newcommand{\eeqa}{\end{eqnarray}}
\newcommand{\be}{\begin{equation}}
\newcommand{\ee}{\end{equation}}
\newcommand{\ba}{\begin{array}} 
\newcommand{\ea}{\end{array}}

\begin{document} 
\title{Partially flavour non-universal $U(1)$ and radiative fermion masses}
\author{Gurucharan Mohanta}
\email{gurucharan@prl.res.in}
\affiliation{Theoretical Physics Division, Physical Research Laboratory, Navarangpura, Ahmedabad-380009, India}
\author{Ketan M. Patel}
\email{ketan.hep@gmail.com}
\affiliation{Theoretical Physics Division, Physical Research Laboratory, Navarangpura, Ahmedabad-380009, India}

\begin{abstract}
We investigate an extension of the Standard Model with a partially flavour non-universal abelian gauge symmetry that enables radiative mass generation for lighter fermions. With flavour-universal charges for the first two generations, gauge-induced loop corrections generate second-generation masses while keeping the first generation massless. Small first-generation masses can arise from subdominant scalar loops within this framework. A single anomaly-free $U(1)$ is sufficient for a realistic model, and the partial universality relaxes the gauge boson mass bound from several thousand TeV to about 200 TeV compared with previous frameworks of this type. We also identify an alignment limit matching the lightest scalar to the observed Higgs and compute its couplings with the fermions, which largely agree with Standard Model values but show testable deviations.
\end{abstract}

\maketitle

\section{Introduction}
\label{sec:intro}
Employing quantum corrections to generate first- and second-generation fermion masses offers one of the most plausible explanations for the observed hierarchies in the masses of quarks and leptons~ \cite{Weinberg:1972ws,Georgi:1972hy,Mohapatra:1974wk,Barr:1978rv}. Realizing this mechanism necessarily requires an extension of the Standard Model (SM), not only through new symmetries to forbid their tree-level masses, but also by introducing new fields to generate them radiatively in a viable manner~\cite{Balakrishna:1987qd,Berezhiani:1992pj,Arkani-Hamed:1996kxn,Barr:2007ma,Dobrescu:2008sz,Weinberg:2020zba}. Extensions with spontaneously broken abelian or non-abelian gauge symmetries, under which the three generations transform non-trivially, most efficiently capture these requirements~\cite{Weinberg:2020zba,Jana:2021tlx,Mohanta:2022seo,Mohanta:2023soi,Bonilla:2023wok,Mohanta:2024wcr,Jana:2024icm}. The underlying symmetry can be utilized to give rise to rank-one mass matrices at tree level, while the new massive gauge boson(s) can induce the required quantum corrections.

The new gauge boson, being essentially flavourful, faces very strong constraints from flavour-changing neutral currents (FCNC). For example, in the case of an abelian gauged symmetry with generic flavour non-universal charges, the gauge boson mass is pushed to at least $10^{4 - 5}$~TeV. This lower limit can be relaxed to $10^3$~TeV by optimising the flavour-dependent charges, as shown in~\cite{Mohanta:2024wcr}. In this case, the first-, second-, and third-generation fermions can have charges $q_1 = 1 - \epsilon$, $q_2 = 1 + \epsilon$, and $q_3 = -2$ under the new $U(1)$. In the limit $\epsilon \to 0$, both the flavour violations involving the first-generation fermions and the quantum corrections to their masses induced by the new gauge boson vanish. While the former helps in evading some of the strongest FCNC constraints, the latter creates a viability issue. Therefore, one requires a small but non-vanishing value of $\epsilon$, disfavoring the possibility $q_1 = q_2$.

Notably, all these extensions also feature an extended scalar sector, required to generate the rank-one structure for fermion mass matrices at leading order. They can also contribute to the light fermion masses through radiative corrections. This work aims to quantify such contributions and assess their implications for effective model building. We show that while it is possible to keep these contributions negligibly small compared to those induced by the gauge boson under reasonable assumptions, they can also play a non-trivial role through their small but non-negligible contributions to fermion masses. The latter opens up an interesting possibility in which the new gauge symmetry need not be completely flavour non-universal. We show that a model with $q_1 = q_2 \neq q_3$ can successfully reproduce both the first- and second-generation masses radiatively.

An immediate consequence of $q_1 = q_2$ is that the FCNC induced by the gauge boson involving first-generation fermions are relatively suppressed in comparison to the previous studies. This allows a factor of five improvement in the lower limit of the new gauge boson mass, reducing the separation between the new physics and electroweak scales. We also evaluate the couplings of the scalars with the fermions beyond the leading order and show how the observed Higgs can be identified with one of the scalars present in the framework.

\section{Hierarchical Fermion Masses}
\label{sec:fm}
We consider the down-type quark sector to demonstrate the essential features of the mechanism. The three generations of the quark doublets and singlets are charged under a local $U(1)_X$ symmetry with charges $(1,1,-2)$. The electroweak doublet and singlet scalars also come in three copies with similar charges under the new abelian symmetry. A pair of vectorlike down-type quarks $D_{L,R}$ is introduced to give rise to rank-1 structure as done earlier in \cite{Mohanta:2024wcr}. The gauge-invariant Yukawa and mass terms for the fermions can be written as
\beqa \label{LY}
-{\cal L}_Y &=& y_{d i}\, \overline{Q}_{L i} H_{d i} D_R + y^\prime_{di}\, \overline{D}_L \eta^{*}_i d_{R i} \nonumber \\
&+& m_D\, \overline{D}_L D_R + {\rm h.c.}\,. \eeqa
Note that $\overline{Q}_{L 1}$ is coupled to $H_{d1}$ and not to $H_{d2}$ as it would have been otherwise allowed by the gauge invariance. This can be arranged by some global symmetry, minimally being a $Z_2$. The need of a pair $H_{d1}$, $H_{d2}$ or $\eta_{1}$, $\eta_{2}$ shall become clear at the end of this section. 

Consequent upon the $U(1)_X$ and electroweak symmetry breaking by the non-vanishing vacuum expectation value (VEV), $\langle \eta^*_{i} \rangle \equiv v_{\eta i}$ and for the electrically neutral component $\langle H^0_{d i}\rangle \equiv v_{d i}$, the tree-level down-type quark mass matrix is given by
\be \label{M0}
\left(M^{(0)}_d \right)_{ij} = -\frac{\mu_{di}\, \mu^\prime_{dj}}{m_D}\,,\ee
with $\mu_{di}=y_{di} v_{di}$ and $\mu^\prime_{di}=y^\prime_{di} v_{\eta i}$, after integrating out the vectorlike pair in the limit $m_D \gg \mu_{d i}^\prime, \mu_{d i}$. The matrix is of rank one, and the state with non-vanishing mass can be identified with the bottom quark:
\be \label{M0_diag}
U_{d L}^{(0) \dagger}\,M_d^{(0)}\,U_{d R}^{(0)} = {\rm Diag.} (0,0,m_b^{(0)})\,.\ee
The presence of a pair of massless states can also be understood in terms of the existence of an accidental global $U(2)_L \times U(2)_R$ symmetry of ${\cal L}_Y$ which remains unbroken in the vacuum \cite{Mohanta:2024wcr}.

The flavour non-universal gauge interactions,
\be \label{LX}
-{\cal L}_X = g_X q_i \left(\overline{d}_{L i} \gamma^\mu d_{L i} + \overline{d}_{R i} \gamma^\mu d_{R i} \right)X_\mu\,, \ee
with $q_i=(1,1,-2)$, does not repsect this $U(2)_L \times U(2)_R$ symmetry in general \cite{Mohanta:2024wcr}. As a result, the quantum corrections to the tree-level mass matrix induced by the above gauge interactions can turn on the masses for some of the otherwise massless states. Indeed, explicit computation leads to \cite{Mohanta:2022seo},
\be \label{dMX}
\left(\delta M_d^{(X)}\right)_{ij} = C\,q_i\,q_j\,\left(M^{(0)}_d\right)_{ij}\,,\ee
at 1-loop, where $C = \frac{g_X^2}{4 \pi^2} (B_0[M_X,m_b^{(0)}]-B_0[M_X,m_D])$ and the loop-integration function is reproduced as Eq. (\ref{B0C0}) in Appendix \ref{app:hff}. As can be seen, the correction is a finite and completely calculable quantity in terms of the gauge coupling and masses of the fermions and gauge boson. 

The nature of the correction, Eq. (\ref{dMX}), is such that it induces non-zero mass for only one of the two massless states of $M^{(0)}_d$ leaving an unbroken $U(1)_L \times U(1)_R$ accidental symmetry \cite{Mohanta:2024wcr}. Moreover, the choice of the charges $q_i=(1,1,-2)$ is such that this symmetry is exact for ${\cal L}_X$ implying that even the higher order loop corrections induced by the gauge interactions will not be able to generate the non-zero mass for this state, see \cite{Mohanta:2024wcr}, for more elaborated discussion on this.

This disastrous situation can be rescued by the loop corrections induced by the scalars. The framework possesses six (nine if we consider three scalars, $H_{ui}$, for the up-type quark sector in the full framework) electrically neutral scalars arising from $H_{di}$ and $\eta_i$. When the underlying gauge symmetry is broken, they can mix and give rise to physical neutral scalar states denoted by $S_a$, see the Appendix \ref{app:scalars}. Subsequently, the Yukawa interactions in Eq. (\ref{LY}), in the physical basis of fermions and scalars, take the form 
\beqa \label{LY_S}
-{\cal L}_Y &\supset& (\tilde{y}_d)_{ia}\, \overline{d}^\prime_{L i} S_a D^\prime_R + (\tilde{y}^\prime_d)_{ia}\, \overline{D}^\prime_{L} S_a d^\prime_{Ri} + \dots \,,\eeqa
at the leading order. Here, 
\beqa \label{y_tilde}
(\tilde{y}_d)_{ia} &=& \sum_j (U_{dL}^{(0)\dagger})_{ij} y_{dj} ({\cal R}_d)_{j a}\,, \nonumber \\
(\tilde{y}^\prime_d)_{ia} &=& \sum_j (U_{dR}^{(0)T})_{ij} y^\prime_{dj} ({\cal R}_\eta)_{j a}\,,\eeqa
where ${\cal R}_{d,\eta}$ are $3 \times 9$ dimensional matrices as defined in Appendix \ref{app:scalars}. The dots in Eq. (\ref{LY_S}) stand for the additional terms suppressed by one or more powers of $v_{\eta i}/m_D$ and/or $v_{d i}/m_D$.

The one-particle-irreducible two-point function at 1-loop, involving the scalars and vectorlike fermion in the loop, evaluated from the interactions in Eq. (\ref{LY_S}), is obtained as
\be \label{sigma_S}
\sigma^{(S)}_{ij} = -\frac{1}{16 \pi^2}\, m_D\, \sum_a (\tilde{y}_d)_{ia} (\tilde{y}^\prime_d)_{ja}\,B_0[m_{S_a},m_D]\,.\ee
In terms of the correction to the down-type quark mass matrix, it can be simplified to
\be \label{dMS}
\left(\delta M_d^{(S)}\right)_{ij} = \left(U_{dL}^{(0)} \sigma^{(S)} U_{d R}^{(0)\dagger} \right)_{ij} \equiv (\epsilon_d)_{ij}\,(M^{(0)}_d)_{ij}\,,\ee
with 
\be \label{eps_ij}
(\epsilon_d)_{ij} = \frac{1}{16 \pi^2} \frac{m_D^2}{v_{di} v_{\eta j}} \sum_a ({\cal R}_d)_{ia} ({\cal R}_\eta)_{ja}\,B_0[m_{S_a},m_D]\,.\ee
Likewise, the gauge boson contribution, the above is also finite and calculable. The divergent part of $(\epsilon_d)_{ij}$ cancels due to the orthogonality of the ${\cal R}$ matrix which implies ${\cal R}_d {\cal R}_\eta^T=0$, see Appendix \ref{app:scalars}.

Using Eqs. (\ref{M0},\ref{dMX},\ref{dMS}), the one-loop corrected effective mass matrix involving three generations of down-type quarks can be written as
\be \label{Md1}
\left(M_d \right)_{ij} = \left(M_d^{(0)} \right)_{ij}\, \left(1 + C q_i q_j + (\epsilon_d)_{ij}\right)\,.\ee
The diagonalisation of the above leads to three massive states,
\be \label{M1_diag}
U_{d L}^{\dagger}\,M_d\,U_{d R} = {\rm Diag.} (m_d,m_s,m_b)\,,\ee
in general. The contribution arising from $(\epsilon_d)_{ij}$ depends on several parameters of the scalar potential. In practice, it is very difficult to fix these parameters observationally. Hence, from the perspective of calculability, this contribution is not as predictive as the gauge boson contribution.

Let us elucidate how $\epsilon_d$ can induce a non-vanishing down quark mass in the present setup. In the limit $(\epsilon_d)_{ij} \to 0$, the fields corresponding to massless state in the physical basis can be obtained in terms of the original states as $d_{L 1}^\prime = \sum_i (U_{dL}^*)_{i1}\,d_{Li}$ and $d_{R 1}^\prime = \sum_i (U_{dR}^*)_{i1}\,d_{Ri}$. The relevant elements of the unitary matrix are given by \cite{Mohanta:2024wcr} 
\beqa \label{ev}
(U_{dL})_{i1} &=& \frac{1}{\sqrt{|\mu_{d1}|^2+|\mu_{d2}|^2}}\,\left(\mu^*_{d2},\,- \mu^*_{d1},\, 0\right)^T\,, \nonumber \\
(U_{dR})_{i1} &=& \frac{1}{\sqrt{|\mu^\prime_{d1}|^2+|\mu^\prime_{d2}|^2}}\,\left(\mu^{\prime}_{d2},\,- \mu^{\prime}_{d1},\, 0\right)^T\,.\eeqa 
The accidental $U(1)_L \times U(1)_R$ symmetry, under which $d^\prime_{L1}$ and $d^\prime_{R1}$ transform non-trivially, is broken by the interactions in Eq. (\ref{LY_S}) if $(\tilde{y}_d)_{1 a}$ and $(\tilde{y}^\prime_d)_{1 a}$ are nonzero. Replacing $U_{dL,dR}^{(0)}$ with $U_{dL,dR}$ in the definition, Eq. (\ref{y_tilde}), and using Eq. (\ref{ev}), we find 
\beqa \label{y_1a}
(\tilde{y}_d)_{1 a} &=& \frac{\mu_{d1} \mu_{d2}}{\sqrt{|\mu_{d1}|^2+|\mu_{d2}|^2}} \left(\frac{({\cal R}_d)_{1a}}{v_{d1}}-\frac{({\cal R}_d)_{2a}}{v_{d2}} \right), \nonumber \\
(\tilde{y}_d^\prime)_{1 a} &=& \frac{\mu^\prime_{d1} \mu^\prime_{d2}}{\sqrt{|\mu^\prime_{d1}|^2+|\mu^\prime_{d2}|^2}} \left(\frac{({\cal R}_\eta)_{1a}}{v_{\eta 1}}-\frac{({\cal R}_\eta)_{2a}}{v_{\eta 2}} \right). \eeqa
Thus, $U(1)_L$ is broken by non-vanishing $\mu_{d1,d2}$ and $({\cal R}_d)_{1a}/v_{d1} \neq ({\cal R}_d)_{2a}/v_{d2}$. Note that the latter necessarily means that the framework must have $H_{d1}$, $H_{d2}$ with non-identical mixing or VEVs. The same applies to $U(1)_R$ and the requirement of a pair of $\eta_{1,2}$. This is the reason why the framework contains three copies of $H_{d i}$ and $\eta_i$ although two (one with $U(1)_X$ charge $1$ and other with $-2$) would have been sufficient to generate the tree-level mass matrix $M_d^{(0)}$ in the desired form, Eq. (\ref{M0}).

Although $\epsilon_d$ depends on several indeterminate parameters, its magnitude can be controlled by imposing reasonable assumptions on the scalar potential. For instance, it can be rendered negligibly small by fine-tuning the couplings between the $H_{di}$ and $\eta_i$ scalars to be vanishingly small. This has been the main assumption in \cite{Mohanta:2022seo,Mohanta:2024wcr} and hence the complete flavour non-universal gauging of flavours was required to make the lightest generation massive. However, with an appropriate choice of the size of couplings between $H_{di}$ and $\eta_i$, small but non-vanishing $\epsilon_d$ can be achieved as outlined in Appendix \ref{app:scalars}.

It is straightforward to generalise the above mechanism to include the up-type quarks and charged leptons. A complete anomaly cancellation also requires three copies of the SM singlet fermions with the same generational $U(1)_X$ charges \cite{Mohanta:2024wcr}. The options to generate viable neutrino masses and leptonic mixing in these kinds of frameworks are already outlined in \cite{Mohanta:2022seo,Mohanta:2024wcr}. Also, it necessitates inclusion of three copies of $H_{ui}$ since the $U(1)_X$ charges prevent $\tilde{H}_{di} = i \sigma_2 H^*_{di}$ to couple to the up-type quarks. Henceforth, we define $v_d^2 = \sum_i v_{di}^2$, $v_u^2 = \sum_i v_{ui}^2$,  $v_\eta^2 = \sum_i v_{\eta i}^2$ with $v_u^2 + v_d^2 = (246\, {\rm GeV})^2 \equiv v^2$.

\section{Flavour Violations}
\label{sec:flv}
The most stringent lower limits on the spectrum of new particles and $U(1)_X$ breaking scale arise from the inherently present FCNC in frameworks like this. The latter are induced at the tree-level itself by $X$, $Z$ and $S_a$ bosons. The first two are quantified through explicit evaluation of the couplings between the fermions and $X$ or $Z$ boson in \cite{Mohanta:2024wcr}.

In the physical basis, the couplings of $X$-boson are given by
\be \label{fv_x}
-{\cal L}_X \supset  g_X \left(Q^d_L\right)_{ij}\, \overline{d}_{L i}^\prime \gamma^\mu d_{L j}^\prime\,X_\mu\,+\,\{L \to R\}, \ee
with $(Q^d_{L,R})_{ij} = \delta_{ij} - 3\, (U^*_{d{L,R}})_{3i}\, (U_{d{L,R}})_{3j}$ for the present model.  It is straightforward to see that $(Q^d_{L,R})_{12} = (Q^d_{L,R})_{13} = 0$ for the $U_{dL,dR}$ given in Eq. (\ref{ev}). Therefore, no tree-level FCNC are induced by the $X$-boson in $1$-$2$ and $1$-$3$ sectors when $(\epsilon_d)_{ij} = 0$. Small $(\epsilon_d)_{ij}$, necessary to induce the first-generation masses, therefore, can lead to non-zero but relatively suppressed flavour violations involving these fermions. The tree-level flavour changing couplings with $Z$ are induced through heavy-light fermion mixing and found to be suppressed by a factor of ${\cal O}(\mu_{di}^2/m_D^2)$. Therefore, $Z$-induced FCNC remain subdominant for $M_X/M_Z \gtrsim m_D^2/v_d^2$. 

The couplings of the physical neutral scalars with the SM fermions can be computed from Eq. (\ref{LY}). In the limit $m_D > v_{\eta i}, v_{di}$, they can be simplified after some straightforward algebra as 
\be \label{fv_S}
-{\cal L}_Y \supset \left(Y^d_a\right)_{ij}\, \overline{d}_{L i}^\prime d_{R j}^\prime\,S_a\,+\,{\rm h.c.}\,, \ee
where, at the leading order, 
\beqa \label{Y_ij}
\left(Y^d_a\right)_{ij} &=& \sum_{k,l} (U_{dL}^*)_{ki} (U_{dR})_{lj}  \nonumber \\
 & &\times  \left(M_d^{(0)}\right)_{kl} \left(\frac{({\cal R}_d)_{ka}}{v_{d k}} + \frac{({\cal R}_\eta)_{la}}{v_{\eta l}} \right)\,,\eeqa
Consequently, the flavour structure of $Y^d_a$ also depends on the parameters of the scalar potential. In the limit $({\cal R}_d)_{1a}/v_{d1} = ({\cal R}_d)_{2a}/v_{d2}$, $({\cal R}_\eta)_{1a}/v_{\eta 1} = ({\cal R}_\eta)_{2a}/v_{\eta 2}$, we verify that $\left(Y^d_a\right)_{1i}=\left(Y^d_a\right)_{i1}=0$ by using Eqs. (\ref{eps_ij},\ref{Md1},\ref{ev})\footnote{In this case, $(\epsilon_d)_{ij}$ are flavour-universal and hence $M_d$ commutes with $M_d^{(0)}+\delta M_d^{(X)}$ allowing one to use Eq. (\ref{ev}) for relevant columns of $U_{dL,dR}$ in (\ref{Y_ij}).}. This is expected as one finds an unbroken $U(1)_L \times U(1)_R$ symmetry in this limit. 

It is noticed that the couplings $Y^d_a$ are generically suppressed by a factor of ${\cal O}(m_b/v_d)$ in comparison to $Q^d_{L,R}$, which are typically of ${\cal O}(1)$. Therefore, the scalar-induced FCNC are to remain sub-dominant, at least in the down-type quark and charged lepton sectors, in comparison to those mediated by $X$-boson for $v_d \sim v$ and similar masses of scalar and vector bosons.

\section{SM-like Higgs and its couplings}
\label{sec:Higgs}
The model should feature a neutral scalar state identifiable with the Higgs boson observed at the LHC, with mass $M_h \simeq 125$~GeV~\cite{ATLAS:2015yey}. Furthermore, its couplings to fermions must align with experimentally measured values. In the present framework, this can be achieved at best by identifying the lightest state $S_1$ as the observed Higgs and assuming an alignment limit:
\be \label{alignment}
({\cal R}_{d,u})_{i1} = \frac{v_{di,ui}}{v}\,~~{\rm for}~i=1,2,3\,.\ee
Along with $v_\eta \gg v_{d,u}$, the above implies 
\be \label{Y1_ij}
Y^d_1 \simeq \frac{1}{v}\, U_{dL}^\dagger M_d^{(0)} U_{dR} \equiv Y^d\,.\ee

Substituting $U_{dL,dR} \to U^{(0)}_{dL,dR}$ and using Eq. (\ref{M0_diag}), one finds at the leading order,
\be \label{Y1_ij_LO}
Y^d \sim {\rm Diag.} (0,0,m_b^{(0)}/v)\,.\ee
Therefore, the alignment limit enforces that only the third generation fermions couple to the SM-like Higgs with coupling proportional to their mass at the leading order.

To find the $S_1$ couplings with lighter fermions, one needs to go beyond the leading order. We compute them at the next-to-leading order (NLO) as described at length in Appendix \ref{app:hff}. The most dominant contribution arises from the vertex correction involving the $X$-boson in the loop. This is estimated as Eq. (\ref{app:dY_ij}) in Appendix \ref{app:hff} using the leading order interactions quantified by Eqs. (\ref{LY_S}), (\ref{fv_x}) and (\ref{fv_S}) and the alignment limit. For $v \ll v_\eta$ and $m_b \ll m_D, M_X$, it can be further simplified to
\beqa \label{dY1_ij}
\delta Y^d_{ij} &\simeq & \frac{g_X^2}{4 \pi^2} \left(Q_L^d\right)_{i3} \left(Q_R^d \right)_{j3} \frac{m^{(0)}_b}{v} \nonumber \\
& & \times\left(B_0[M_X,m_b^{(0)}]-B_0[M_X,m_D]\right)\,,\eeqa
which is finite and calculable.

Again, using $U_{dL,dR} \to U^{(0)}_{dL,dR}$ as the leading order approximation and from Eq. (\ref{dMX}), one can approximate the above $\delta Y^d$ to
\beqa \label{dY1_ij_LO}
\delta Y^d & \sim & \frac{1}{v}\,U_{dL}^{(0) \dagger}\,\delta M_d^{(X)}\,U_{dR}^{(0)}\,,\eeqa
which, in conjunction with Eq. (\ref{Y1_ij_LO}), implies
\be \label{Y1_ij_NLO}
Y^d + \delta Y^d \sim {\rm Diag.} (0,m_s/v,m_b/v)\,.\ee
Once the sub-dominant scalar-induced corrections are also taken into account, the couplings with the first generation fermions are also expected to take small but non-vanishing values. The above result is very close to what one expects from the SM Higgs, which has flavour-conserving couplings with fermions proportional to the latter's masses. Therefore, Eq. (\ref{alignment}) provides a phenomenologically useful limit in which the lightest of the neutral scalars can resemble the SM Higgs.

\section{Results}
\label{sec:results}
As discussed earlier, finite and non-universal $\epsilon_{d,u}$ are crucial for generating first-generation fermion masses, but they also induce direct flavour-violating couplings involving the first generation, which are tightly constrained. We perform a detailed numerical analysis to quantify this interplay.

We first perform $\chi^2$ fits to evaluate the viability of Eq. (\ref{Md1}) and similar expressions for the up-type quark and the charged lepton matrices, in reproducing the charged fermion masses and quark mixing matrix. The procedure, along with the parameter counting and various perturbativity constraints we put on the parameters of the model, is described in our previous work \cite{Mohanta:2024wcr}. To be specific, we choose non-zero $(\epsilon_{u,d})_{11} = (\epsilon_{u,d})_{22}$ and set the remaining elements to zero. This choice is sufficient to induce the non-vanishing masses for the first-generation fermions. For the reference values of the charged fermion masses, we use the extrapolated data at the renormalisation scale $\mu = 10^5$ GeV given in \cite{Huang:2020hdv}. For the quark mixing angles and CP phase, we use the values from \cite{ParticleDataGroup:2024cfk}. These values are summarised in Table \ref{tab:observables} in Appendix \ref{app:Numerical}.

Once the fit is obtained, we use the fitted parameters of the model to estimate the most relevant observables quantifying the flavour violations, which include meson-antimeson transitions for the neutral mesons $K$, $B_d$, $B_s$, and $D$ and various lepton flavour violating decays summarised in \cite{Mohanta:2024wcr}. The values of relevant Wilson coefficients are derived from the fitted parameters at $\mu = 10^5$ GeV and then run down to the applicable scale of the underlying process and compared with the existing constraints. Further details of explicit computation and presently allowed ranges or limits are described in  \cite{Mohanta:2022seo,Mohanta:2024wcr}. We consider only the $X$-boson induced contributions in estimating these observables, as they are the most dominant and relevant within the parameter space obtained, as discussed in section~\ref{sec:flv}.

\begin{figure}[t!]
\centering
\subfigure{\includegraphics[width=0.24\textwidth]{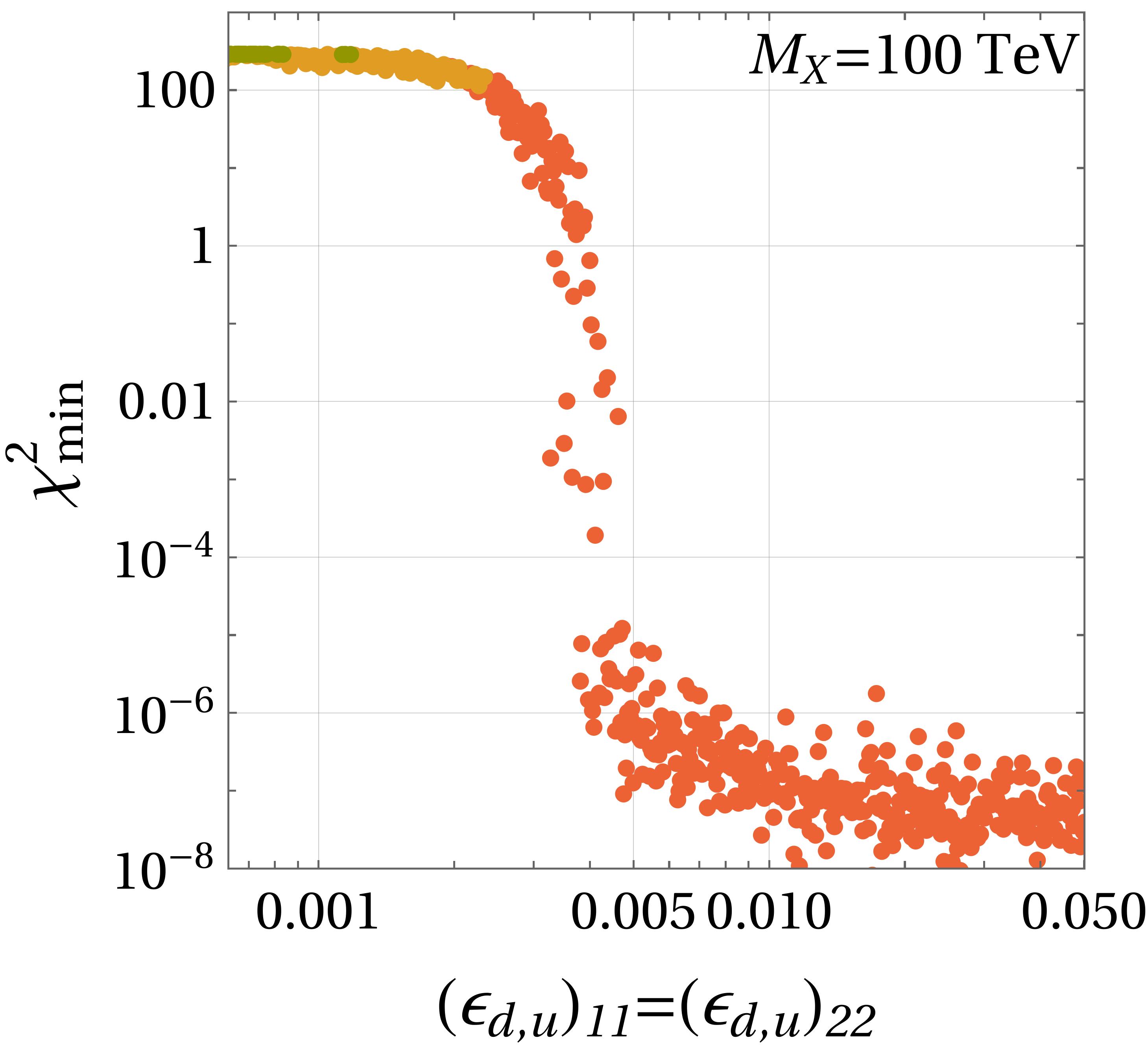}}\hspace*{-0.2cm}
\subfigure{\includegraphics[width=0.24\textwidth]{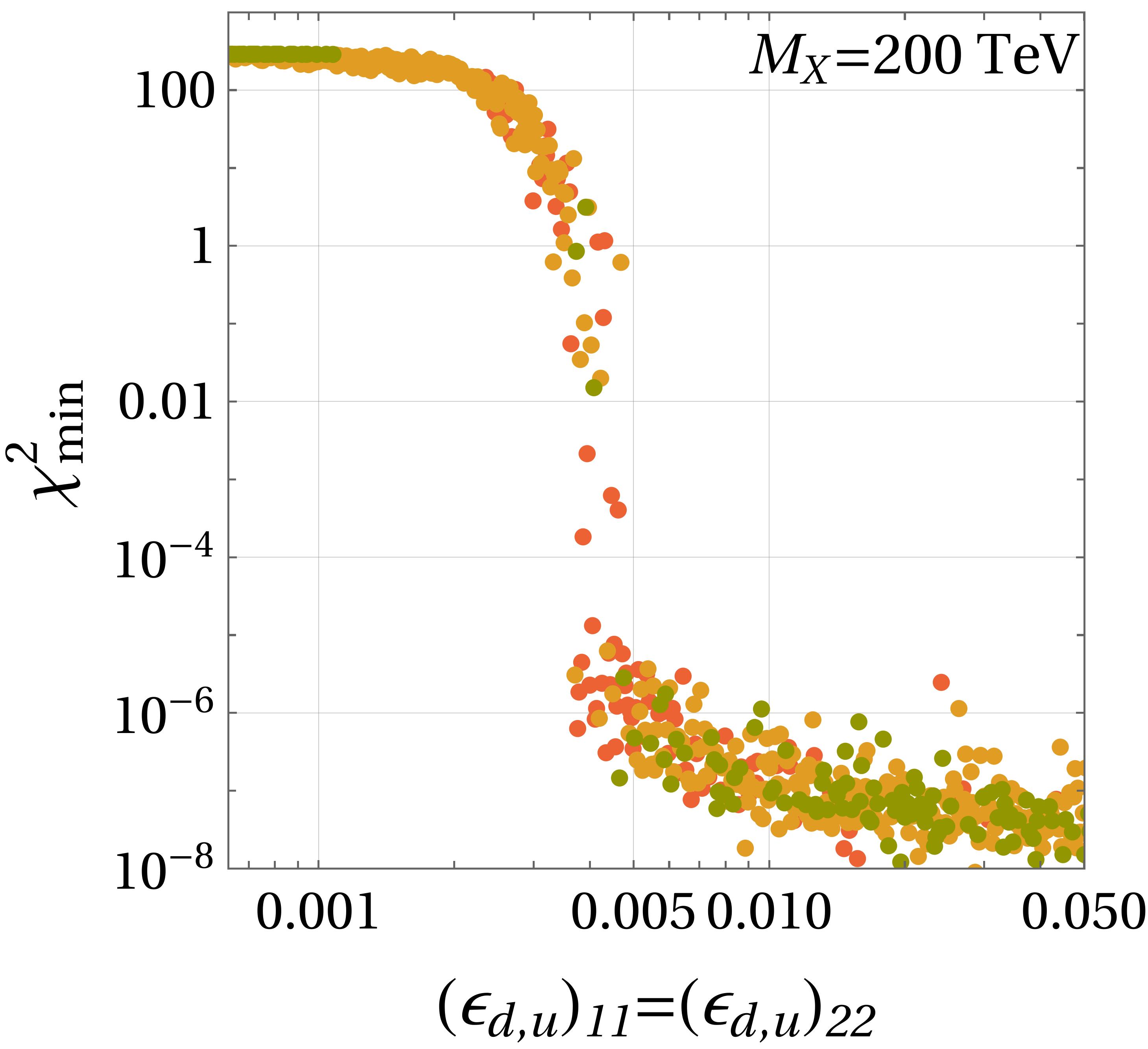}}
\subfigure{\includegraphics[width=0.24\textwidth]{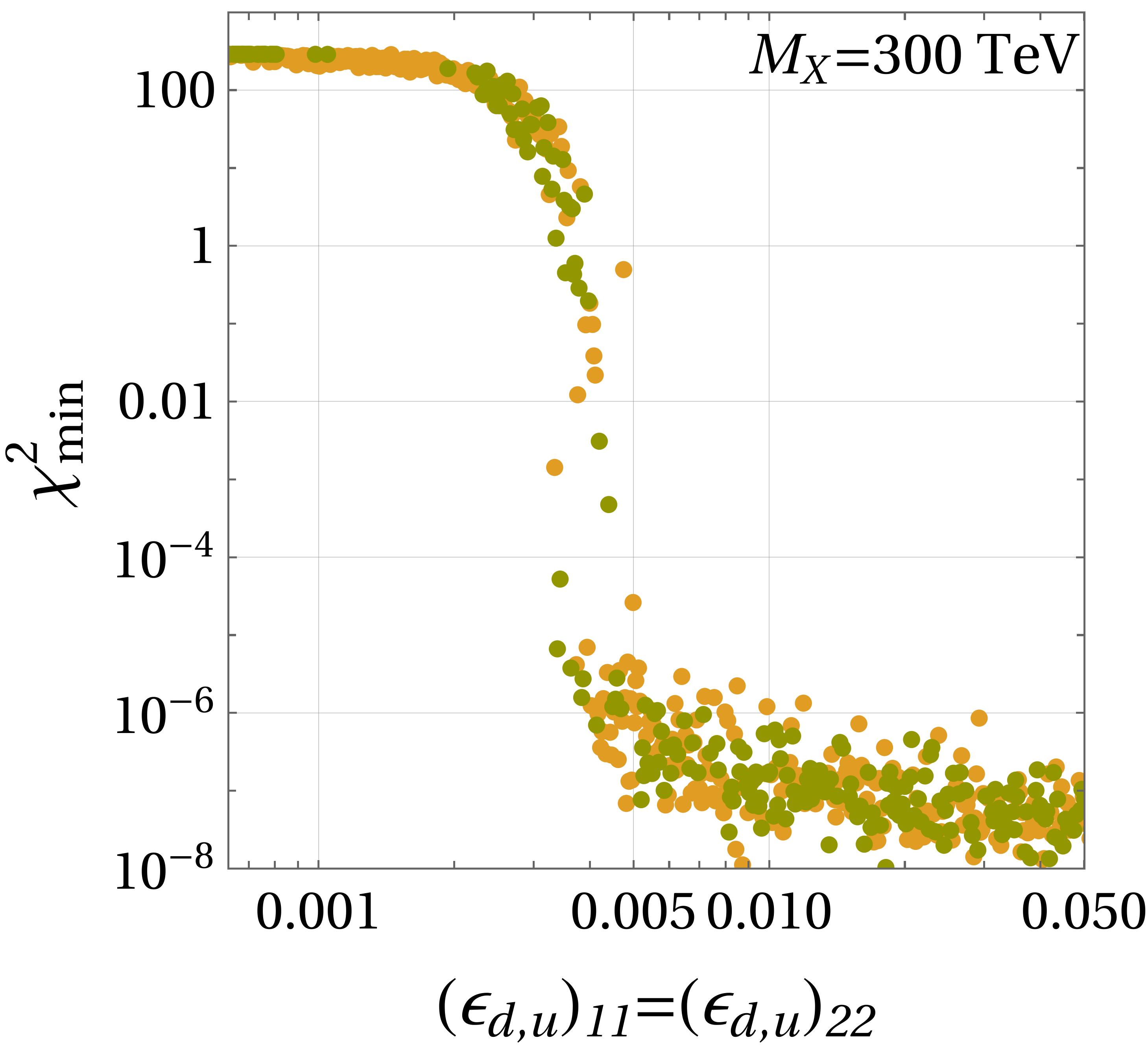}}\hspace*{-0.2cm}
\subfigure{\includegraphics[width=0.24\textwidth]{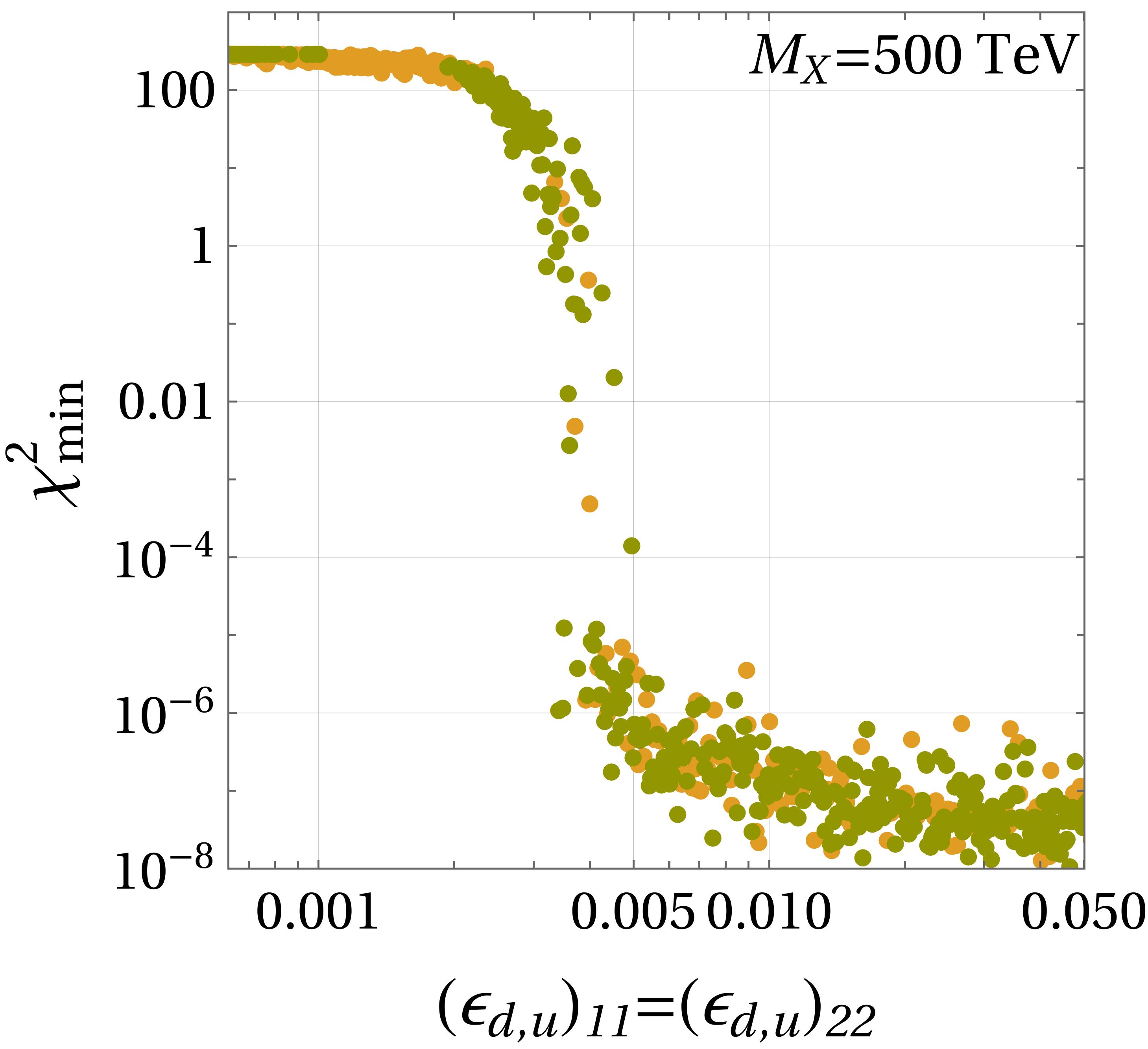}}
\caption{Minimized $\chi^2$ as function of $(\epsilon_{u,d})_{11}=(\epsilon_{u,d})_{22}$ for four sample values of $M_X$. Green points satisfy \emph{all} FCNC constraints discussed in the text. Yellow points pass FCNC tests involving second–third generation transitions but are excluded by constraints from the flavour violations in $1$-$2$ or $1$-$3$ sectors. Red points are disfavoured by FCNC limits in the $2$-$3$ sector.}
\label{fig:fg1}
\end{figure}

The results are displayed in Fig.~\ref{fig:fg1}, which shows the minimised $\chi^2$ values as functions of the non-zero elements of $\epsilon_{u,d}$ for four representative $M_X$ values. The information about flavour violations in different sectors is encoded in the colours. As $\epsilon_{u,d}$ approaches zero, the first generations become massless, resulting in large $\chi^2_{\rm min}$. All the constraints from flavour violations are satisfied for these points as in the  $\epsilon_{u,d} \to 0$ limit, direct flavour violation in $1$-$2$ and $1$-$3$ sector becomes negligible while the FCNC effects in $2$-$3$ sector are also aptly suppressed for $M_X \gtrsim 50$ TeV. Switching on sizeable $\epsilon_{u,d}$ improves $\chi^2_{\rm min}$, mainly because the first-generation fermions obtain the required masses, but also introduce significant FCNCs mainly in $1$-$2$ sector. The constraints from the latter require relatively higher $M_X$ to obtain the viable charged fermion masses and quark mixing. Although we have carried out the above analysis for a particular ansatz for $\epsilon_{u,d}$ matrices, we find similar results for other choices of  $\epsilon_{u,d}$ as long as they are not completely universal.

It is found that $\chi^2_{\rm min} \leq 1$, and consistency with FCNC constraints requires $M_X \geq 200$ TeV. This is significantly lower than the bound $M_X \gtrsim 10^5$ TeV obtained in our earlier work on fully flavour non-universal gauge symmetries for radiative fermion mass generation \cite{Mohanta:2022seo}. It is also a factor of 5 smaller than the typical upper limit found in a scenario with gauged flavour symmetry with mildly broken universality in the $1$-$2$ sector \cite{Mohanta:2024wcr}. This suggests that the extension with an abelian symmetry carrying charges $(1,1,-2)$, along with relatively small but finite scalar-sector corrections, can offer a viable framework for radiative fermion masses with a new physics scale in the few hundred TeV range. 

We give the values of input parameters in Table \ref{tab:input} in Appendix \ref{app:Numerical} for two benchmark best-fit points corresponding to $M_X=200$ and $300$ TeV and optimum $\epsilon_{u,d}$ in each case such that $\chi^2_{\rm min} < 1$.  All the observables listed in Table \ref{tab:observables} are reproduced with negligible deviations from their central values. We also list the estimated values of various FCNC observables considered in the present study for these benchmark points in Table \ref{tab:meson_WC}. It is seen that the lower limit on $M_X$ is still mainly governed by the constraints coming from the $K$-$\overline{K}$ transition.

\begin{figure}[t!]
\centering
\subfigure{\includegraphics[width=0.22\textwidth]{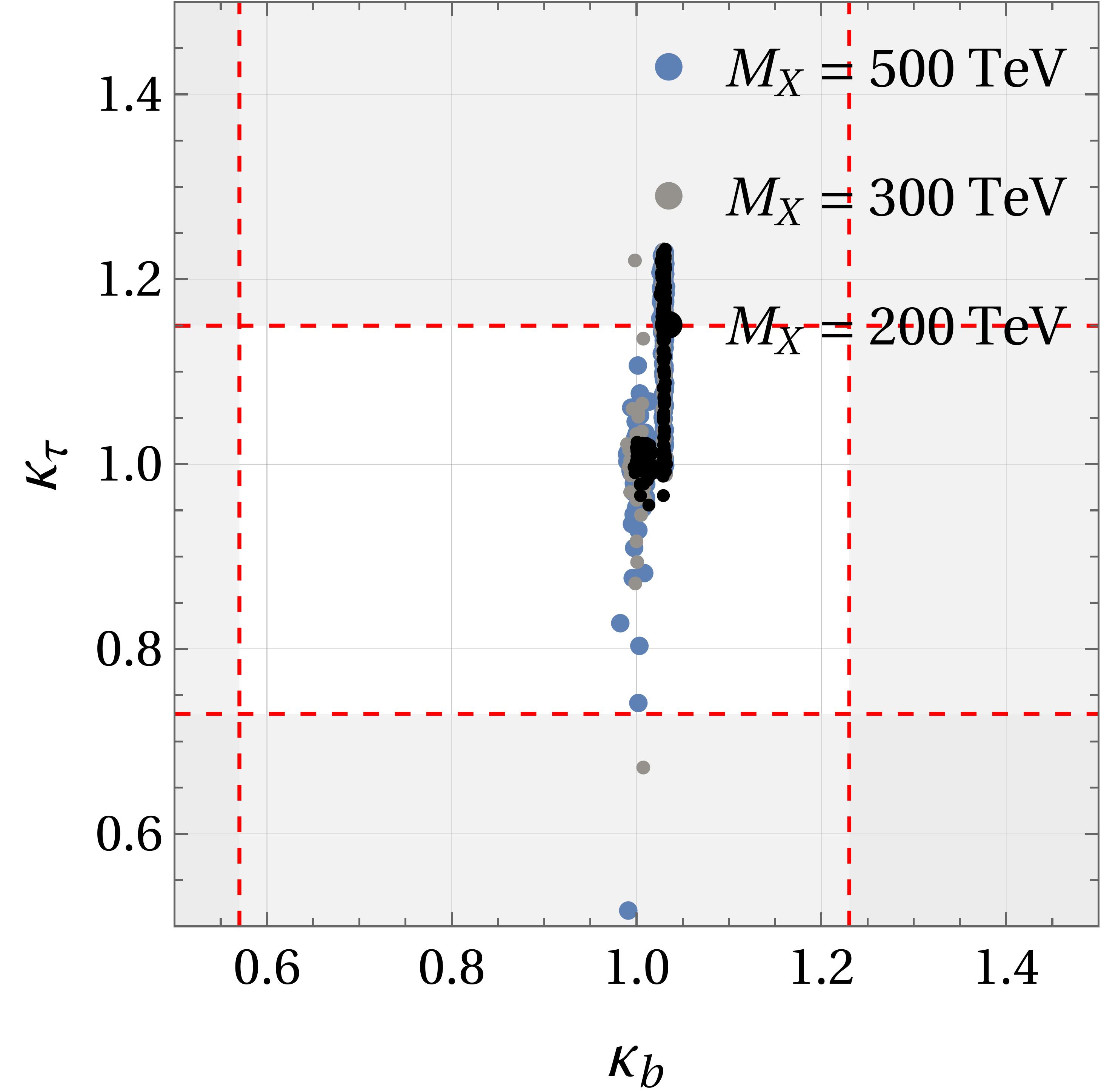}}
\subfigure{\includegraphics[width=0.24\textwidth]{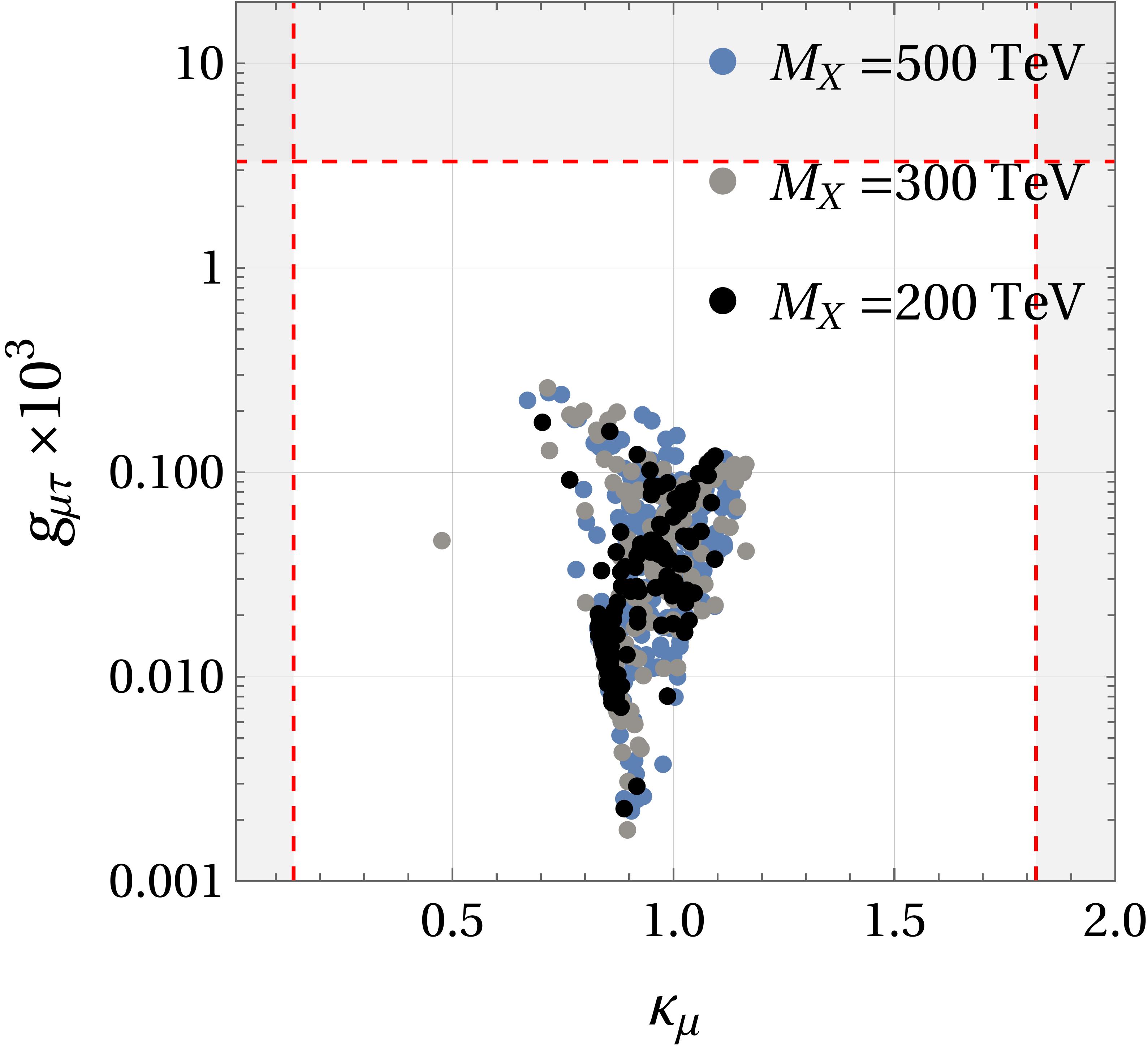}}
\caption{Predictions for the relevant couplings of  SM-like Higgs $S_1$ with fermions in the alignment limit for all the viable solutions. $\kappa_b = |Y^d_{33}|/(m_b/v)$, $\kappa_\tau = |Y^e_{33}|/(m_\tau/v)$ and $\kappa_\mu = |Y^e_{22}|/(m_\mu/v)$ are coupling strengths normalised to their SM values and the corresponding experimental ranges are taken from \cite{ATLAS:2022vkf}. $g_{\mu \tau}=\sqrt{|Y^e_{23}|^2+|Y^e_{32}|^2}$ and its limit is taken from \cite{ATLAS:2023mvd}. The unshaded regions between the dashed lines stand for the allowed range at $3\sigma$.}
\label{fg2}
\end{figure}

As discussed in section~\ref{sec:Higgs}, the state $S_1$ in the alignment limit corresponds to the SM-like Higgs boson. Its couplings to fermions may offer valuable insights into the underlying theoretical framework. We estimate the couplings with $b\,\bar{b}$, $\tau^+\,\tau^-$, $\mu^+\,\mu^-$ and $\mu^\pm\,\tau^\mp$, which are presently measured with a reasonably good precision, for all the viable solutions with $\chi^2_{\rm min} \leq 1$. The estimation is carried out using Eqs. (\ref{Y1_ij}) and (\ref{dY1_ij}), which capture leading order as well as the most dominant NLO contributions. The results are displayed in Fig. \ref{fg2} and a comparison is made with the present experimental constraints.
Notably, a large number of the predicted values lie within the experimentally allowed range, despite exhibiting significant deviations from their SM values. In particular, the couplings to charged leptons show sizeable deviations, and more precise measurements could serve as a sensitive probe of the alignment limit within this framework. It is also worth to note that the size of deviation is almost independent of the values of $M_X$.

\section{Conclusions}
\label{sec:cncl}
We show that the SM, extended with a partially flavour non-universal abelian gauge symmetry, can effectively accommodate a radiative mass generation mechanism for the lighter fermions. The most effective implementation arises when the first two generations carry flavour-universal charges. As a result, gauge boson-induced corrections generate loop-suppressed masses for the second generation while keeping the first generation massless. The latter can acquire small masses through relatively suppressed scalar-induced radiative corrections, which are, anyway, present as an inherent feature of this class of frameworks.

There are two significant improvements over previously considered gauged extensions of this type. First, a single $U(1)$ symmetry with simple and anomaly-free charges is sufficient to construct a complete and realistic model. Second, the universality of charges for the first two generations leads to suppressed flavour transition rates involving first-generation fermions. This, in turn, lowers the lower bound on the gauge boson mass from several thousand TeV to approximately 200 TeV.

We also propose an alignment limit in which the lightest of the neutral scalars can be identified with the Higgs boson observed at the LHC, and compute its couplings to fermions in this framework for the first time. These couplings, when evaluated at next-to-leading order, are shown to largely agree with their SM expectations, while still exhibiting significant deviations that can serve as a powerful probe of such frameworks even when the scale of new physics lies beyond the reach of direct searches.

\begin{acknowledgments}
We thank Jacky Kumar for useful discussions on the latest FCNC constraints. This work is supported by the Department of Space (DOS), Government of India.
\end{acknowledgments}

\bibliography{references}

\pagebreak
\appendix 
\onecolumngrid\
\section{Scalar sector}
\label{app:scalars}
We provide an elaborate discussion on the scalar sector of the theory and comment on the size of scalar-induced radiative corrections to the fermion masses in this Appendix. The scalar sector contains three copies of electroweak doublets $H_{ui}$, $H_{di}$ and singlets $\eta_i$. The three copies are charged as $(1,1,-2)$ under the $U(1)_X$ as discussed in section \ref{sec:fm}. Additionally, to get the Yukawa interactions in Eq. (\ref{LY}) in the desired form, a $Z_2$ symmetry can be considered under which $H_{u1}$, $H_{d1}$ and $\eta_1$ are odd while the rest of the scalars are even.

The most general renormalisable scalar potential invariant under the full gauge symmetry of the model and $Z_2$ is given by,
\beqa \label{potential}
V &=& \frac{1}{2} m_{u i}^2\, H_{u i}^\dagger H_{u i}\, +\, \frac{1}{2}m_{d i}^2\, H_{d i}^\dagger H_{d i}\, +\, \frac{1}{2}m_{\eta i}^2\, \eta_{i}^\dagger \eta_{i}\,+\, \left\{ (m_{\eta})_{p}\, \eta_p \eta_p \eta_3 + {\rm h.c.}\right\}\, \nonumber \\
&+&\, \left\{(m_{ud\eta})_{1p}\, \eta_p H_{u p} H_{d 3}\,+\, (m_{ud\eta})_{2p}\, \eta_p H_{u 3} H_{d p}\,+\,(m_{ud\eta})_{3p}\, \eta_3 H_{u p} H_{d p}\,+\,{\rm h.c.} \right\} \\
& + &  \frac{1}{4}(\lambda_u)_{ij}\, H_{u i}^\dagger H_{u i} H_{u j}^\dagger H_{u j}\,+\, \frac{1}{4}(\lambda_d)_{ij}\, H_{d i}^\dagger H_{d i} H_{d j}^\dagger H_{d j}\,+\, \frac{1}{4}(\lambda_\eta)_{ij}\, \eta_{i}^\dagger \eta_{i} \eta_{j}^\dagger \eta_{j}\,+\,\frac{1}{4}(\lambda_{ud})_{ij}\, H_{u i}^\dagger H_{u i} H_{d j}^\dagger H_{d j}\, \nonumber \\
& + & \frac{1}{4}(\lambda_{u\eta})_{ij}\, H_{u i}^\dagger H_{u i} \eta_j^\dagger \eta_j\,+\, \frac{1}{4}(\lambda_{d \eta})_{ij}\, H_{d i}^\dagger H_{d i} \eta_{j}^\dagger \eta_{j}\,+\,\frac{1}{2} (\tilde{\lambda}_{ud})_{ij}\, H_{u i}^\dagger H_{u j} H_{d j}^\dagger H_{d i}\,+\, \frac{1}{2}(\tilde{\lambda}_{u\eta})_{ij}\, H_{u i}^\dagger H_{u j} \eta_j^\dagger \eta_i\, \nonumber \\
& + & \frac{1}{2}(\tilde{\lambda}_{d \eta})_{ij}\, H_{d i}^\dagger H_{d j} \eta_{j}^\dagger \eta_{i}\, +\,\left\{(\lambda_{ud\eta})_{ij}\, \eta_i^\dagger H_{ui} \eta_j^\dagger H_{dj} + (\tilde{\lambda}_{ud\eta})_{ij}\, \eta_i^\dagger H_{uj} \eta_j^\dagger H_{di} + {\rm h.c.}\right\}\,,\eeqa
where $i=1,2,3$ and $p=1,2$. In the above, $\lambda_{u,d,\eta}$ are real and symmetric $3\times 3$ matrices. $\lambda_{ud}$, $\lambda_{u\eta}$, $\lambda_{d\eta}$, $\tilde{\lambda}_{ud}$, $\tilde{\lambda}_{u\eta}$, $\tilde{\lambda}_{d\eta}$ are hermitian matrices and one can set $(\tilde{\lambda}_{ud})_{ii} = (\tilde{\lambda}_{u\eta})_{ii} = (\tilde{\lambda}_{d \eta})_{ii} = 0$ without loss of generality. 

Further restrictions can be imposed on $V$ if additional symmetries are considered. For example, an additional $Z_2$ under which either $H_{ui}$ or $H_{di}$ are assumed to be odd eliminates $m_{ud\eta},\lambda_{ud\eta}$ and $\tilde{\lambda}_{ud\eta}$ entirely. Similarly, assigning $\eta_i$ a $Z_2$-odd charge sets trilinear coupling matrix $m_\eta$ to zero. The $U_R$ or $D_R$, $E_R$ can be made odd accordingly, and this symmetry is softly broken by the masses of vectorlike fermions. Even with such additional restrictions, the scalar sector contains a considerable number of indeterministic parameters. Dependency on these parameters enters into the flavour sector through next-to-leading order corrections.

To parametrise and estimate the sizes of these contributions, let us consider the spectrum of the electrically neutral scalars. There are nine scalars to be denoted by $\tilde{S} = (H_{ui}^0,H_{di}^0,\eta^*_{i})^T$, where the latter are fields after substituting their respective values in the vacuum. All these fields mix through the mass terms that can be parametrised in general as,
\be \label{M_Sa_def}
V \supset \frac{1}{2}\,({\cal M}^2_S)_{ab}\, \tilde{S}_a \tilde{S}_b\,,\ee
where $a=1,...,9$, and 
\be \label{M_Sa}
{\cal M}^2_S = \left(\ba{ccc} m_{u u}^2 & m^2_{ud} &  m^2_{u\eta} \\ (m^2_{ud})^T & m_{d d}^2 & m^2_{d\eta}\\ (m^2_{u\eta})^T & (m^2_{d\eta})^T & m^2_{\eta \eta} \ea \right)\,.\ee

Each element in ${\cal M}^2_S$ represents a $3 \times 3$ block and they can arise in the following way. Assuming that the VEVs of $\eta_i \sim {\cal O}(M_X)$ are sufficiently larger than the electroweak scale, the diagonal and off-diagonal elements of $m^2_{\eta \eta}$ receive dominant contributions from the former,
\beqa \label{m_etaeta}
\left(m^2_{\eta \eta}\right)_{ii} &=& m_{\eta i}^2+2 v_{\eta i}^2 (\lambda_{\eta})_{ii} + \sum_{j} v_{\eta j}^2 (\lambda_{\eta})_{ij} + {\cal O}\left(v^2 \right),\nonumber \\
\left(m^2_{\eta \eta}\right)_{ij} &\underset{i \neq j}{=}& 2 v_{\eta i} v_{\eta j} (\lambda_{\eta})_{ij}+ {\cal O}\left(v^2 \right).\eeqa
Similarly, one finds
\beqa \label{m_dd}
\left(m^2_{dd}\right)_{ii} &=& m_{di}^2 + \frac{1}{2} \sum_{j} v_{\eta j}^2 (\lambda_{d\eta})_{ij} + {\cal O}\left(v^2 \right),\nonumber \\
\left(m^2_{dd}\right)_{ij} &\underset{i \neq j}{=}& v_{\eta i} v_{\eta j} \left((\tilde{\lambda}_{d\eta})_{ij}+(\tilde{\lambda}_{d\eta})_{ji}\right)+ {\cal O}\left(v^2 \right),\eeqa
for $m^2_{dd}$ and a similar expression for $m^2_{uu}$ with appropriate changes. Therefore, all the diagonal blocks are of similar sizes in general. An arrangement like $m^2_{uu},m^2_{dd} \ll m_{\eta \eta}^2$ typically requires $\lambda_{d\eta} \sim \tilde{\lambda}_{d \eta} \sim \lambda_{u\eta} \sim \tilde{\lambda}_{u \eta} \lesssim {\cal O}(v^2/M_X^2)$ and $m^2_{di},m^2_{ui} \ll m^2_{\eta \eta}$. 

Unlike the above, the off-diagonal blocks $m^2_{u\eta}$, $m^2_{d\eta}$ and $m^2_{ud}$ get generated only when electroweak symmetry is broken for the vanishing $\lambda_{ud\eta}$, $\tilde{\lambda}_{ud\eta}$ and trilinear couplings. One finds in this case,  
\beqa \label{m_deta}
\left(m^2_{d\eta}\right)_{ii} &=& 2 v_{d i} v_{\eta i} (\lambda_{d\eta})_{ii} +  \sum_{j\neq i} v_{d j} v_{\eta j} \left((\tilde{\lambda}_{d\eta})_{ij}+(\tilde{\lambda}_{d\eta})_{ji}\right),\nonumber \\
\left(m^2_{d\eta}\right)_{ij} &\underset{i \neq j}{=}&2 v_{d i} v_{\eta j} (\lambda_{d \eta})_{ij} +  v_{d j} v_{\eta i} \left((\tilde{\lambda}_{d\eta})_{ij}+(\tilde{\lambda}_{d\eta})_{ji}\right),\eeqa
and a similar expression for $m^2_{u\eta}$. Moreover,
\beqa \label{m_ud}
\left(m^2_{ud}\right)_{ii} &=& 2 v_{u i} v_{d i} (\lambda_{ud})_{ii} +  \sum_{j\neq i} v_{u j} v_{d j} \left((\tilde{\lambda}_{ud})_{ij}+(\tilde{\lambda}_{ud})_{ji}\right),\nonumber \\
\left(m^2_{ud}\right)_{ij} &\underset{i \neq j}{=}& 2 v_{u i} v_{d j} (\lambda_{ud})_{ij} +  v_{u j} v_{d i} \left((\tilde{\lambda}_{ud})_{ij}+(\tilde{\lambda}_{ud})_{ji}\right).\eeqa
If all the non-vanishing quartic couplings are of ${\cal O}(1)$, then $m^2_{\eta \eta}, m^2_{uu}, m^2_{dd} \sim {\cal O}(M_X^2)$, $m^2_{d\eta}, m^2_{u\eta}\sim {\cal O}(v M_X)$ and $m^2_{ud} \sim{\cal O}(v^2)$ follow from the above expressions if no fine-tuned arrangements between different contributions are assumed. Similarly, for $ \lambda_{d \eta}, \tilde{\lambda}_{d \eta}, \lambda_{u\eta},\tilde{\lambda}_{u \eta} \sim {\cal O}(v^2/M_X^2)$, one finds $m^2_{\eta \eta} \sim {\cal O}(M_X^2)$, $m^2_{uu}, m^2_{dd} , m^2_{ud} \sim {\cal O}(v^2)$, and  $m^2_{d\eta}, m^2_{u\eta}\sim {\cal O}(v^3/M_X)$.

The neutral scalar fields $\tilde{S}_a$ can be brought into the physical basis $S_a$ by transformation, $\tilde{S}_a = {\cal R}_{ab} S_b$, such that ${\cal R}^T {\cal M}_S^2 {\cal R}={\rm Diag.}(m_{S_1}^2,...,m_{S_9}^2)$. ${\cal R}$ is a $9 \times 9$ complex orthogonal matrix in general, which can be conveniently parametrised as 
\be \label{R9}
{\cal R} = \left(\ba{c} {\cal R}_{u} \\ {\cal R}_{d} \\ {\cal R}_{\eta} \ea \right)=\left(\ba{ccc} R_{uu} & R_{ud} &  R_{u\eta} \\ R_{du} & R_{d d} & R_{d\eta}\\ R_{\eta u} & R_{\eta d} & R_{\eta \eta} \ea \right)\,.\ee
As noted earlier, $m_{u\eta}^2,m_{d \eta}^2 \ll m_{\eta \eta}^2$ for $M_X \gg v_{u,d}$. In this case, the relevant $3 \times 3$ blocks of ${\cal R}$ can be approximated as,
\beqa \label{R_approx}
R_{u \eta} &\simeq & m^2_{u \eta} (m^2_{\eta \eta})^{-1} R_{\eta \eta}\,, ~~~~~~~~~~
R_{d \eta} \simeq  m^2_{d \eta} (m^2_{\eta \eta})^{-1} R_{\eta \eta}\,, \nonumber \\
R_{\eta u} &\simeq & - \left(m^2_{u \eta} (m^2_{\eta \eta})^{-1} \right)^T R_{uu}\,, ~~
R_{\eta d} \simeq  - \left(m^2_{d \eta} (m^2_{\eta \eta})^{-1} \right)^T R_{dd}\,. \eeqa
Following the discussion presented in the previous paragraph and above expressions, the magnitudes of $R_{u \eta}, R_{d \eta}, R_{\eta u}, R_{\eta d}$ ranges from ${\cal O}(v^3/M_X^3)$ to ${\cal O}(v/M_X)$, depending on the strengths of the couplings characterising interactions between $H_{di,ui}$ and $\eta_i$.

The above also determines the size of scalar-induced 1-loop correction to fermion mass matrix quantified as $(\epsilon_d)_{ij}$ in Eq. (\ref{eps_ij}). Estimating $m_D \sim M_X$, $v_{di} \sim v$ and $v_{\eta i} \sim M_X$, one typically finds 
\be \label{eps_size}
 \left|(\epsilon_d)_{ij}\right| \sim \frac{1}{2 \pi^2} \times \begin{cases}
 {\cal O}\left(\frac{v^2}{M_X^2}\right),~\text{for $\lambda_{d \eta}, \tilde{\lambda}_{d \eta} \sim {\cal O}\left(\frac{v^2}{M_X^2}\right)$}\\
 {\cal O}(1),~\text{for $\lambda_{d \eta}, \tilde{\lambda}_{d \eta} \sim {\cal O}(1)$}. \end{cases} \ee
Arbitrarily small $(\epsilon_d)_{ij}$ can also be arranged by assuming sufficiently small values of the underlying quartic couplings. It is important to note that the scalar potential does not exhibit an enhanced symmetry in the limit $\lambda_{d \eta}, \tilde{\lambda}_{d \eta}, \lambda_{u \eta}, \tilde{\lambda}_{u \eta} \to 0$. Consequently, the smallness of these couplings is not technically natural and must be regarded as a result of fine-tuning. This situation is analogous to the gauge hierarchy problem, which already afflicts models of this kind due to the significant separation between the electroweak and $U(1)_X$ breaking scales.

\section{Evaluation of Higgs-Fermion-Fermion couplings at NLO}
\label{app:hff}
The lightest of the neutral scalars, namely $S_1$, which is identified as the SM like Higgs, couples to only the third generation fermions at the leading order as shown in section \ref{sec:Higgs} in the alignment limit. In this Appendix, we estimate the next-to-leading order correction to these couplings. 

The most dominant correction to the Yukawa vertex arises from the one-loop diagrams involving a chiral, a vectorlike and an $X$-boson in the loop. There is a direct coupling between $d^\prime_i$, $D^\prime$ and $S_1$ through Eq. (\ref{LY_S}). Moreover, Eq. (\ref{LX}), when written in the mass basis, gives rise to coupling between $d^\prime_i$, $D^\prime$ and $X$-boson of ${\cal O}(\mu^\prime_{di}/m_D)$ or  ${\cal O}(\mu_{di}/m_D)$. The correction to $Y^d$ induced by these diagrams is obtained as 
\beqa \label{dY2_def}
-i \left(\delta Y^d\right)_{ij} & = & - \frac{g_X^2}{4 \pi^4} \sum_{k=1}^3 \Big\{ \left( \left(Q_L^d\right)_{ik} Y^d_{k4}  \left(Q_R^d \right)_{j4} + \left(Q_L^d\right)_{i4} Y^d_{4k}  \left(Q_R^d \right)_{jk} \right) \int d^4q\, \frac{q^2}{(q^2-m_k^2) (q^2-m_D^2) (q^2-M_X^2)} \nonumber \\
& +&\left( \left(Q_L^d\right)_{ik} Y^{d*}_{4k}  \left(Q_R^d \right)_{j4} + \left(Q_L^d\right)_{i4} Y^{d*}_{k4}  \left(Q_R^d \right)_{jk} \right)\,m_k m_D\,\int d^4q\, \frac{1}{(q^2-m_k^2) (q^2-m_D^2) (q^2-M_X^2)} \Big\}\,.\eeqa
Here,
\be \label{Q_4i}
\left(Q^d_{R}\right)_{i4} = \frac{1}{m_D} \sum_{j,k} \left(Q^d_{R}\right)_{ij} (U_{dR})_{jk} \mu^\prime_k\,, ~~\left(Q^d_{L}\right)_{4i}= \frac{1}{m_D} \sum_{j,k} \mu^*_k (U_{dL})_{kj} \left(Q^d_L\right)_{ki}\,.\ee
are obtained from Eq. (\ref{LX}). Using the alignment limit, Eq. (\ref{alignment}), and similar one for ${\cal R}_\eta$, the Yukawa couplings can be similified as,
\beqa \label{Y_4i}
Y^d_{i 4} &=& \left(\tilde{y}_d\right)_{i1}=\sum_j (U_{dL}^\dagger)_{ij} y_{dj} ({\cal R}_d)_{j1} = \frac{1}{v} \sum_j (U_{dL}^\dagger)_{ij} \mu_{dj}\,,\nonumber\\
Y^d_{4i} &=& \left(\tilde{y}^\prime_d\right)_{i1}=\sum_j (U_{dR}^T)_{ij} y^\prime_{dj} ({\cal R}_\eta)_{j1} = \frac{1}{v_\eta} \sum_j (U_{dR}^T)_{ij} \mu^\prime_{dj}\,.\eeqa

Evaluating the integrals, one finds that the terms in the second line of Eq. (\ref{dY2_def}) are suppressed by a factor of ${\cal O}(m_b/m_D)$ in comparison to those in the first line. Further, substituting Eqs. (\ref{Q_4i},\ref{Y_4i}) in Eq. (\ref{dY2_def}) and using the leading order result, Eq. (\ref{Y1_ij_LO}), the above contribution can be simplified to
\beqa \label{dY2_res}
\delta Y^d_{ij} & = -& \frac{g_X^2}{4 \pi^2} \frac{m_b^{(0)}}{v} \left(Q_L^d\right)_{i3}  \left(Q_R^d \right)_{j3} \Big\{B_0[M_X,m_D] + (m_b^{(0)})^2\, C_0[M_X,m_D,m_b^{(0)}] \Big\}\,,\eeqa
where we have also neglected the terms suppressed by ${\cal O} (v/v_\eta)$. The integration functions are given by
\beqa \label{B0C0}
B_0[m_1,m_2] & = & 1- \frac{m_1^2 \ln \frac{m_1^2}{\mu^2}-m_2^2 \ln \frac{m_2^2}{\mu^2}}{m_1^2-m_2^2}\,, \nonumber \\
C_0[m_0,m_1,m_2] &=& -\frac{1}{m_0^2}\frac{-t_1 \ln{t_1} + t_1t_2 \ln{t_1}+t_2\ln{t_2}-t_1t_2\ln{t_2}}{(-1+t_1)(t_1-t_2)(-1+t_2)}\,,\eeqa
with $t_p =m_p^2/m_0^2$.

Considering the possibility of two vectorlike fermions in the loop, it can be noted that they do not have coupling with $S_1$ at the leading order. Moreover, such a diagram shall have both $(Q^d_R)_{i4}$ and  $(Q^d_L)_{i4}$ at vertices giving rise to a suppressed contribution. Therefore, these contributions can be neglected. For the light fermions in the loop, the correction to $Y^d$ is evaluated as,
\beqa \label{dY1_def}
-i \left(\delta Y^d\right)_{ij} & = & - \frac{g_X^2}{4 \pi^4} \sum_{k,l=1}^3 \Big\{\left(Q_L^d\right)_{ik} Y^d_{kl}  \left(Q_R^d \right)_{jl}\,\int d^4q\, \frac{q^2}{(q^2-m_k^2) (q^2-m_l^2) (q^2-M_X^2)} \nonumber \\
& +& \left(Q_L^d\right)_{ik} Y^{d *}_{lk}  \left(Q_R^d \right)_{jl}\,m_k m_l\,\int d^4q\, \frac{1}{(q^2-m_k^2) (q^2-m_l^2) (q^2-M_X^2)} \Big\}\,.\eeqa
Again, using the leading order result, Eq. (\ref{Y1_ij_LO}), and the alignment limit, we find
\beqa \label{dY1_res}
\delta Y^d_{ij} & = & \frac{g_X^2}{4 \pi^2} \frac{m_b^{(0)}}{v} \left(Q_L^d\right)_{i3}  \left(Q_R^d \right)_{j3} \Big\{B_0[M_X,m_b^{(0)}] + 2(m_b^{(0)})^2\, C_0[M_X,m_b^{(0)},m_b^{(0)}] \Big\}\,.\eeqa

The Higgs-Fermion-Fermion couplings also receive corrections at 1-loop from scalars in the loop. The most dominant among these arise from a diagram involving one SM and one vectorlike fermion in the loop. These interactions are parametrised in Eqs. (\ref{LY_S},\ref{fv_S}). Computing them explicitly, we find 
\beqa \label{dY3_def}
-i \left(\delta Y^d\right)_{ij} & = & - \frac{1}{16 \pi^4} \sum_{a=1}^9 \sum_{k=1}^3 \Big\{ \left( (\tilde{y}_d)_{ia} (\tilde{y}_d)^*_{k1} \left(Y^d_a\right)_{kj}  + \left(Y^d_a\right)_{ik} (\tilde{y}_d^\prime)_{k1}^* (\tilde{y}^\prime)_{ja} \right) \int d^4q\, \frac{q^2}{(q^2-m_k^2) (q^2-m_D^2) (q^2-M_{S_a}^2)} \nonumber \\
& +&\left( (\tilde{y}_d)_{ia} (\tilde{y}^\prime_d)_{k1} \left(Y^d_a\right)_{kj}  + \left(Y^d_a\right)_{ik} (\tilde{y}_d)_{k1} (\tilde{y}^\prime)_{ja} \right)\,m_k m_D\,\int d^4q\, \frac{1}{(q^2-m_k^2) (q^2-m_D^2) (q^2-M_{S_a}^2)} \Big\}\,.\eeqa
It can be noted that this contribution is relatively small due to the small Yukawa couplings. Indeed, since $\tilde{y}_d \sim {\cal O}(\mu_d/v_d)$, $\tilde{y}^\prime_d \sim {\cal O}(\mu_d^\prime/v_\eta)$ and $Y^d_a \sim {\cal O}(m_b/v_d)$, one typically finds a suppression with respect to Eq. (\ref{dY2_res}) by a factor of ${\cal O}(\mu_d^2/v_d^2)$ or ${\cal O}(\mu_d^{\prime 2}/v_\eta^2)$. These are small, in particular for the down-type quarks and the charged leptons, for $v_d \sim v$. This can be explicitly seen from the optimised values of the input parameters reported in Table \ref{tab:input} for two benchmark points. 

Combining all these results, the most dominant NLO contribution to the Higgs-Fermion-Fermion couplings is estimated from Eqs. (\ref{dY2_res},\ref{dY1_res}) as,
\beqa \label{app:dY_ij}
\delta Y^d_{ij} & \approx & \frac{g_X^2}{4 \pi^2} \left(Q_L^d\right)_{i3} \left(Q_R^d \right)_{j3} \frac{m^{(0)}_b}{v} \times \Bigl\{B_0[M_X,m_b^{(0)}]-B_0[M_X,m_D]  \nonumber \\
&+& (m_b^{(0)})^2\, \left( 2 C_0[M_X,m_b^{(0)},m_b^{(0)}] - C_0[M_X,m_D,m_b^{(0)}] \right)  \Bigr\}\,.\eeqa
It can be further noticed that the last two terms remain negligibly small for $m_D,M_X \gg m_b$.

\section{Benchmark solutions and FCNC predictions}
\label{app:Numerical}
We provide here the supplementary material related to numerical analysis and results discussed in section \ref{sec:results}. The values of the charged fermion masses and quark mixing parameters used as reference are summarised in Table \ref{tab:observables}.
\begin{table}[t]
\begin{center}
\begin{tabular}{cccccc} 
\hline
\hline
~~Observable~~&~~Value~~&~~Observable~~&~~Value~~\\
 \hline
$m_u$ & $0.86\pm 0.15$ MeV & $m_e$ & $0.499 \pm 0.049$ MeV \\
$m_c$ & $0.435 \pm 0.013$ GeV & $m_\mu$ &  $0.105\pm 0.0105$ GeV \\
$m_t$ & $123.77\pm 0.85$ GeV & $m_\tau$ & $1.784 \pm 0.1784$ GeV\\
$m_d$ & $1.88 \pm 0.13$ MeV & $|V_{us}|$ & $0.22501 \pm 0.00068$ \\
$m_s$ & $0.03747 \pm 0.00326$ GeV & $|V_{cb}|$ & $0.04183 \pm 0.00079$  \\
$m_b$ & $1.908 \pm 0.021$ GeV & $|V_{ub}|$ & $0.003732\pm 0.00009$ \\
 &  & $J_{\rm CP}$ & $(3.12 \pm 0.13)\times 10^{-5}$ \\
\hline
\hline
\end{tabular}
\end{center}
\caption{The values of the charged fermion masses and quark mixing parameters used in the numerical analysis.}
\label{tab:observables}
\end{table}
The values of charged fermion masses we use are at the renormalisation scale $\mu=10^5$ GeV and are obtained from \cite{Huang:2020hdv}. For the quark masses, we use the extrapolated uncertainties as reported in \cite{Huang:2020hdv}. However, for the charged leptons, we consider a conservative $\pm 10\%$ uncertainties from their central values. The values of quark mixing and CP phase are taken from \cite{ParticleDataGroup:2024cfk}. The definition of the $\chi^2$ function and optimisation procedure are identical to the ones used and described in detail in \cite{Mohanta:2022seo,Mohanta:2024wcr}.

The complete set of model parameters and their counting is similar to one given in \cite{Mohanta:2024wcr}. Analogous to $\mu_{di}$ and $\mu^\prime_{di}$, we have $\mu_{ui} = y_{ui} v_{ui}$, $\mu^\prime_{ui} = y^\prime_{ui} v_{\eta i}$ for the up-type quarks and $\mu_{ei} = y_{ei} v_{di}$, $\mu^\prime_{ei} = y^\prime_{ei} v_{\eta i}$ for the charged leptons. The masses of vectorlike down-type, up-type quarks and charged leptons are denoted by $m_D$, $m_U$ and $m_E$, respectively. The scalar induced corrections are characterised in terms of only one parameter, $(\epsilon_{u,d})_{11} = (\epsilon_{u,d})_{22}$. These, along with the $X$-boson mass, form a total of 25 real parameters after rotating away the unphysical phases \cite{Mohanta:2024wcr}. We also set $g_X = 1/2$ for definiteness while performing the numerical analysis. The values of these parameters at the minimum of $\chi^2$ for two benchmark points are reproduced in Table \ref{tab:input}.
\begin{table}[!t]
\begin{center}
\begin{tabular}{cccc} 
\hline
\hline
~~Parameters~~&~~Solution 1 (S1)~~   &~~Solution 2 (S2)~~\\
 \hline
$M_X$     & $2 \times10^5$       & $3 \times 10^5$ \\
$(\epsilon_{d,u})_{11,22}$    &  $3.7325\times10^{-3}$             &   $3.3884\times10^{-3}$  \\
$m_U$      & $5.4600 \times 10^{4}$  & $9.2825\times10^{4}$\\
$m_D$      & $9.2520 \times 10^{5}$  & $9.4291\times10^{5}$ \\
$m_E$      &  $1.0694 \times 10^{7}$ & $2.5291\times10^{7}$ \\
\hline
$\mu_{u1}$ &  $-5.2086 \times10^{1}$  & $-4.7203\times10^{1}$\\
$\mu_{u2}$ & $5.4320\times10^{1}$    & $-4.4077 \times10^{1}$\\
$\mu_{u3}$ & $6.2367\times10^{1}$    & $-3.8024 \times10^{1}$\\
\hline
$\mu^\prime_{u1}$ &  $4.9027\times 10^{4}$  & $3.7090 \times10^{2}$\\
$\mu^\prime_{u2}$ & $-1.700 \times 10^2$     & $1.0721\times10^{5}$\\
$\mu^\prime_{u3}$ & $4.8598\times 10^4$      & $1.0949\times10^{5}$\\
\hline
$\mu_{d1}$   & $4.3560  -i ~ 8.4183\times 10^{-2}$ & $3.2690 +i ~ 2.3134\times 10^{-1} $\\
$\mu_{d2}$   &  $-4.6413 +i~ 2.6735\times 10^{-2}$  & $3.0029 +i~ 2.6361 \times 10^{-1}$ \\
$\mu_{d3}$   &  $-6.0477$                           & $2.8559$ \\
\hline
$\mu^\prime_{d1}$ &  $7.9624\times 10^{4}$  & $-2.9559\times 10^{5}$\\
$\mu^\prime_{d2}$ &  $1.8033\times 10^{5}$  & $-1.2869\times 10^{5}$\\
$\mu^\prime_{d3}$ &  $-4.4047\times 10^{4}$  & $1.0534\times 10^{5}$\\
\hline
$\mu_{e1}$   &  $5.0668\times 10^{1}$  & $2.8424 \times 10^{1}$\\
$\mu_{e2}$   &  $9.4104\times 10^{1}$ & $-1.0975 \times 10^{2}$\\
$\mu_{e3}$   &  $1.0962\times 10^{2}$ & $2.6772  \times 10^{1}$\\
\hline
$\mu^\prime_{e1}$ &  $1.8772\times 10^{4}$  & $-7.0054\times 10^{4}$\\
$\mu^\prime_{e2}$ &  $2.9820\times 10^{4}$  & $-2.7508\times 10^{5}$\\
$\mu^\prime_{e3}$ &  $-1.0956\times 10^{5}$  & $-2.7313\times 10^{5}$\\
\hline
\hline
\end{tabular}
\end{center}
\caption{Optimised values of the explicit model parameters obtained for two sample solutions. The values of all the dimension-full parameters are in GeV.}
\label{tab:input}
\end{table}
The solutions are chosen for $M_X=200$ and $300$ TeV and for the smallest $(\epsilon_{d,u})_{11,22}$ leading to $\chi^2_{\rm} \lesssim 1$. As a result of the latter, they reproduce all the observables listed in Table \ref{tab:observables} within $1\sigma$ from their central values. 

\begin{table}[!ht]
\begin{center}
\begin{tabular}{cc}
\hline
\hline
   Quark Sector  & \begin{tabular}{cccc} 

~~Observable  ~~&~~Allowed range~~       &~~S1~~                      &~~S2~~\\
 \hline
Re$C_K^1$ & $[-9.6,9.6]\times 10^{-13}$         &  $  1.2\times 10^{-15}$   & $ -2.0 \times 10^{-15}$   \\
Re$\tilde{C}_K^1$ & $[-9.6,9.6]\times 10^{-13}$ &  $  2.7\times 10^{-16}$   & $ -1.4 \times 10^{-16}$   \\
Re$C_K^4$ & $[-3.6,3.6]\times 10^{-15}$         &  $ -3.0 \times 10^{-15}$   & $ 2.8 \times 10^{-15}$  \\
Re$C_K^5$ & $[-1.0,1.0]\times 10^{-14}$         &  $ -2.5 \times 10^{-15}$   & $ 2.4 \times 10^{-15}$ \\
Im$C_K^1$ & $[-9.6,9.6]\times 10^{-13}$         &  $ -1.4 \times 10^{-28}$   & $ 1.0 \times 10^{-27}$  \\
Im$\tilde{C}_K^1$ & $[-9.6,9.6]\times 10^{-13}$ &  $ 3.2 \times 10^{-29}$   & $ -6.8 \times 10^{-29}$  \\
Im${C}_K^4$ & $[-1.8,0.9]\times 10^{-17}$       &  $ -2.4 \times 10^{-31}$   & $ 6.0 \times 10^{-31}$  \\
Im$C_K^5$ & $[-1.0,1.0]\times 10^{-14}$         &  $ -2.0 \times 10^{-31}$   & $  5.1\times 10^{-31}$   \\
\hline
$|C_{B_d}^1|$ & $<2.3\times 10^{-11}$         &  $ 1.0 \times 10^{-15}$   & $ 8.2 \times 10^{-16}$   \\
$|\tilde{C}_{B_d}^1|$ & $<2.3\times 10^{-11}$ &  $ 1.5 \times 10^{-17}$   & $ 1.5 \times 10^{-17}$   \\
$|C_{B_d}^4|$ &  $<2.1\times 10^{-13}$        &  $ 3.1 \times 10^{-16}$   & $ 2.9 \times 10^{-16}$   \\
$|C_{B_d}^5|$ & $<6.0\times 10^{-13}$         &  $ 5.2 \times 10^{-16}$   & $ 4.8 \times 10^{-16}$   \\
\hline
$|C_{B_s}^1|$ & $< 1.1 \times 10^{-9}$          &  $ 1.2 \times 10^{-11}$   & $ 4.2 \times 10^{-12}$   \\
$|\tilde{C}_{B_s}^1|$ & $< 1.1 \times 10^{-9}$  &  $ 2.2 \times 10^{-12}$   & $ 1.8 \times 10^{-12}$   \\
$|C_{B_s}^4|$ & $< 1.6 \times 10^{-11}$         &  $ 1.3 \times 10^{-11}$   & $ 7.0 \times 10^{-12}$   \\
$|C_{B_s}^5|$ & $< 4.5 \times 10^{-11}$         &  $  2.2\times 10^{-11}$   & $ 1.2 \times 10^{-11}$   \\
\hline
$|C_D^1|$       & $<7.2 \times 10^{-13}$    &  $ 6.6 \times 10^{-13}$   & $ 4.6 \times 10^{-13}$   \\
$|\tilde{C}_D^1|$ & $<7.2 \times 10^{-13}$  &  $ 1.6 \times 10^{-18}$   & $ 5.9 \times 10^{-19}$   \\
$|C_D^4|$      & $<4.8\times 10^{-14}$      &  $ 4.0 \times 10^{-15}$   & $  2.0\times 10^{-15}$   \\
 $|C_D^5|$     & $<4.8 \times 10^{-13}$     & $ 4.6 \times 10^{-15}$    & $ 2.3 \times 10^{-15}$  \\
\end{tabular} \\
\hline
 Lepton Sector &  \begin{tabular}{c@{\hspace{20pt}}c@{\hspace{15pt}}c@{\hspace{10pt}}c}
${\rm BR}[\mu \to e]$      & $< 7.0 \times 10^{-13}$ & $ 1.0 \times 10^{-16}$ & $ 9.2\times 10^{-16}$ \\
\hline
${\rm BR}[\mu \to 3e]$     & $< 1.0 \times 10^{-12}$ & $ 5.7 \times 10^{-19}$ & $5.2 \times 10^{-18}$\\
${\rm BR}[\tau \to 3\mu]$  & $< 1.8 \times 10^{-8}$  & $ 1.2 \times 10^{-14}$ &$ 2.3 \times 10^{-14}$ \\
${\rm BR}[\tau \to 3 e]$   & $< 2.7 \times 10^{-8}$  & $ 2.2 \times 10^{-19}$ & $ 8.2 \times 10^{-20}$\\
\hline
${\rm BR}[\mu \to e \gamma]$    & $< 4.2 \times 10^{-13}$ & $ 6.4 \times 10^{-19}$  & $ 5.1 \times 10^{-19}$ \\
${\rm BR}[\tau \to \mu \gamma]$ &  $< 4.4 \times 10^{-8}$ &  $ 1.6 \times 10^{-16}$ & $ 2.0 \times 10^{-17}$ \\
${\rm BR}[\tau \to e \gamma]$   &  $< 3.3 \times 10^{-8}$ & $ 8.2\times 10^{-22}$   &  $6.5 \times 10^{-23}$\\

\end{tabular}\\
\hline
\hline
\end{tabular}
\end{center}
\caption{Various FCNC observables estimated for the two example solutions along with their present experimental limits. The values of Wilson coefficients are in ${\rm GeV}^{-2}$.}
\label{tab:meson_WC}
\end{table}
The most relevant observables quantifying the flavour violations in both the quark and lepton sectors obtained for the two example solutions are tabulated in Table \ref{tab:meson_WC}. As described in section \ref{sec:results}, these include neutral meson-antimeson transitions involving $K$, $B_d$, $B_s$ and $D$ mesons and lepton flavour violating processes. The estimation of all these quantities is carried out following the strategy described at length in  \cite{Mohanta:2024wcr}. As the present limits on these observables, we use the bounds at $95\%$ confidence level from \cite{UTfit:2007eik,Aebischer:2020dsw} for the Wilson coefficients. For the branching ratios of the charged lepton flavour violating processes, we use limits at $90 \%$ confidence level from \cite{Calibbi:2017uvl}.

\end{document}